\documentclass[final,5p]{elsarticle}
\usepackage{graphicx}
\usepackage{amsmath}

\begin{document}

\begin{frontmatter}
\title{Ordering in rolled-up single-walled ferromagnetic nanomembranes}
\author[PWr]{Andrzej Janutka}
\address[PWr]{Department of Theoretical Physics, Faculty of Fundamental Problems of Technology, Wroclaw University of Technology, 50-370 Wroc{\l}aw, Poland}

\begin{abstract}
Magnetization of soft-ferromagnetic nano- and microtubes of nanometer-thin 
 walls (a single-widening rolled-up nanomembranes) is theoretically studied 
 using analytical and numerical approaches including different stress-induced
 anisotropies. Within the analytical study, we consider magnetostatic effects
 qualitatively, with an effective anisotropy, while they are fully treated
 in the micromagnetic simulations (limited to the tubes of submicrometer
 diameters however). Basic types of the periodic ordering have been established
 and their presence in nanotubes of polycrystalline Permalloy and cobalt 
 has been verified within the simulations. The domain structure is basically
 determined by a material-deposition-induced helical stress or a cooling-induced
 axial stress via the volume magnetostriction while it is influenced by
 the distribution of magnetic charges as well. Also, it is dependent on the initial
 state of the magnetization process.
\end{abstract}

\begin{keyword}
ferromagnetic nanotube, ferromagnetic microtube, stress-driven anisotropy,
magnetic domain structure, analytical micromagnetism, micromagnetic simulations
\end{keyword}
\end{frontmatter}

\footnotemark{E-mail address: Andrzej.Janutka@pwr.edu.pl}
\section{Introduction}

Ordering in a small ferromagnetic tube with very thin wall compared to its radius is
 difficult to anticipate since that structure shares features of the thin film and magnetic wire
 relevant to the magnetostatics (a radial anisotropy of the hard-axis type and easy long axis
 of the tube) with a strong stress-driven anisotropy dependent on fabrication conditions.
 Because of complexity of the anisotropy, there are many metastable states of the magnetization,
 thus, the ordering is sensitive to initial conditions and external factors. On the other hand,
 the tube is a very important geometry among magnetic nano- and micro-systems since
 tubular coverings enable modifications of magneto-transport; GMI effect, and magneto-optical
 properties of wires and fibers to be utilized for sensing applications \cite{buz06,vaz13,sed00}.
 Also, magnetic microtube can serve as a sensing pipe for magnetic nanofluids \cite{mon11}.
 
Techniques of manufacturing single-crystalline and polycrystalline micro- and nanotubes
 of magnetic materials include electrochemical and chemical routes \cite{vaz13,sui04,nie05,ye12}.
 Highly efficient production methods are developed for magnetic microtubes of a thin wall.
 They are produced with sputtering in the form of microwire or microfiber coverings or rolled-up
 membranes of nanometer thicknesses \cite{bor15,str14}. 
 Note that an outer shell of the amorphous glass-coated magnetic microwire can be considered as a tube
 as well, albeit it strongly interacts with the inner core of the wire, (the glass-coated magnetic microwire
 is a single-phase system with a core-shell type magnetic ordering \cite{chi96}), and its wall is quite
 thick relative to the wire radius \cite{ant97,tor11}. However, upon glass removal, the outer shell becomes
 very thin while the thickness of a domain wall (DW) that separates it from the inner core increases \cite{ova09,zha13}.
 This is accompanied by a reorganization of the domain structure and influences the GMI characteristics \cite{kur11}.
 With regard to functionalized wires and fibers, there is a need for modeling the dynamics 
 of the thin-wall tube magnetization. The first step to do in order to formulate an effective
 model is to understand dominant mechanisms responsible for ordering in the nano- and
 micro-tubes without external influences. 
  
In microtubes of very small wall-thickness to radius ratio, the longitudinal easy-axis
 anisotropy of the magnetostatic origin is weaker than in tubes of a thick wall
 or in wires. Thus, influenced by the stress-driven anisotropy, the domain magnetization
 can strongly deviate from the long-axis direction even in very elongated systems. 
 Moreover, the magnetostatically-induced hard-axis anisotropy (the hard axis is normal
 to the tube surface) is strong, which facilitates in-the-wall ordering (excluding
 singularities; vortex and antivortex cores) independent of 
 the stress direction and sign of the volume magnetostriction.
 Despite the shape anisotropy is not well defined
 in the system with an inhomogeneous magnetization, any efficient analytical approach
 to establishing equilibrium states of the tube requires introducing such an effective 
 anisotropy into the model. Full micromagnetic simulations are necessary to verify
 the validity of such a simplification to the nanotubes while they are not any efficient
 alternative to the analytical evaluations of the microtube characteristics at present.
 It is because, simulating microtubes requires too large computational resources.
 
The purpose of the present study is to identify basic equilibrium states of 
 thin-wall microtubes and nanotubes including longitudinal, transverse,
 and helical anisotropies. We compare static analytical
 and numerical solutions to the Landau-Lifshitz-Gilbert equation for nanotubes,
 modeling the magnetostatics effect
 with an effective anisotropy or performing full micromagnetic calculations, respectively.
 The evaluations are focused on the polycrystalline tubes of the most popular magnetic 
 materials; Co and Py tubes. When exclude the crystalline anisotropy effect, important 
 differences in ordering of these two materials follow from different saturation magnetizations.
 The influence of other factors (origin of the internal stress, initial state
 of the magnetization) on the formation of the magnetic texture is discussed as well.
 
In section II, a model of the thin-wall nano- and microtube is formulated, its analytical  
 solutions are pointed out. Section III is devoted to presenting results of micromagnetic
 simulations of the process of tube ordering. Conclusions are collected in section IV.  

\section{Model}

In our analytical approach to study the magnetization of a polycrystalline or amorphous tube, 
 the LLG equation in 3D is included in the form
\begin{eqnarray}
-\frac{\partial{\bf m}}{\partial t}&=&\frac{J}{M_{s}}{\bf m}\times\Delta{\bf m}
+\frac{\beta_{1}}{M_{s}}({\bf m}\cdot\hat{i}){\bf m}\times\hat{i}
\nonumber\\
&&-\frac{\beta_{2}}{M_{s}}
\frac{[{\bf m}\cdot(0,y,z)]{\bf m}\times(0,y,z)}{(y^{2}+z^{2})}
\nonumber\\
&&+\frac{\beta_{3}}{M_{s}}
\frac{[{\bf m}\cdot(0,-z,y)]{\bf m}\times(0,-z,y)}{(y^{2}+z^{2})}
\nonumber\\
&&+\frac{\beta_{4}}{M_{s}}({\bf m}\cdot{\bf a}){\bf m}\times{\bf a}
-\frac{\alpha}{M_{s}}{\bf m}\times\frac{\partial{\bf m}}{\partial t}.
\label{LLG}
\end{eqnarray}
Here, $\hat{i}\equiv[1,0,0]$, (the wire is directed along the $x$ axis), $M_{s}=|{\bf m}|$
 represents the saturation magnetization, $J$ denotes the exchange
 constant, ($J\equiv2\gamma A_{ex}/M_{s}$; $A_{ex}$ is called an exchange stiffness
 while $\gamma$ a gyromagnetic factor), $\beta_{1}$,
 $\beta_{2}$, and $\beta_{3}$ determine the strength of effective axial, radial,
 and circumferential anisotropies, respectively. An additional anisotropy in the tube wall
 is included with the $\beta_{4}$ constant. In the relevant term of the torque, 
 ${\bf a}$ is a combination of $\hat{i}$ and $(0,-z,y)/\sqrt{y^{2}+z^{2}}$ vectors
 and $|{\bf a}|=1$.

In order to establish main contributions to the anisotropy constants, utilizing analogies
 to the tubes, we adapt elements of the theory of elasticity of the amorphous
 and polycrystalline glass-coated microwires which is well developed \cite{chi96}.
 According to this theory, the internal stress can be created at two production stages;
 the solidification of the magnetic material and its cooling that is a much slower process.
 The solidification of the magnetic microwire develops in the radial direction.
 In the surface layer of the microwire that can be considered as a tube, it produces equal
 to each other axial and circumferential components of the stress while the relevant radial
 stress is negligibly small. This kind of stress follows from a homogeneous shrinking of 
 the inside surface of the tube compared to the outside surface and we call
 it a "solidification stress", 
 (in the body of the "rapidly-solidified" wire, the relevant stress is created layer by layer).
 In the presence of that stress, in the tubes made of amorphous or polycrystalline materials,
 the magnetostriction that is of the volume type only (isotropic) is expected to equally
 contribute to the $\beta_{1}$ and $\beta_{3}$ constants of the anisotropy.

However, the tube manufacturing is a different process from the wire production in general.
 Typically, the magnetic tubes are formed via rolling-up planar magnetic (created with
 sputtering or evaporation) layers or via direct sputtering on a surface of cylindrical wires.
 Those methods of the material deposition are accompanied by another "solidification stress"
 and a resulting easy direction in the magnetic layer that is parallel or perpendicular
 to the sputtering (evaporation) plane usually. While the rapid-solidification
 stress is not expected to be strong in a very thin film, the directed sputtering can result
 in the creation of a significant anisotropy relevant to $\beta_{4}$ constant \cite{dij01,che07}.  
 Note that such an anisotropy can be weakened or completely removed via annealing. 
 That "helical" anisotropy in a magnetic tube has been modeled previously in \cite{uso14}. 
 
Another type of the stress can dominate in multi-layered tubes. 
 Since the thermal expansion is isotropic within
 the cross-section of the double-layer tube, for a sufficiently long tube, 
 a difference in the thermal expansion coefficients of the magnetic and non-magnetic
 layers results in equal to each other radial
 and circumferential stresses as well as in a much higher axial stress which are induced
 during the slow cooling process. Therefore, in the amorphous or polycrystalline tubes,
 the "cooling stress" contributions to the constants of the radial and circumferential
 anisotropies are equal. 

Denoting the magnetostatic, solidification,
 and cooling contributions to the anisotropy constants with the relevant indices;
 $\beta_{i}=\beta_{i}^{(ms)}+\beta_{i}^{(solid)}+\beta_{i}^{(cool)}$, ($i=1,2,3$), we establish
 $\beta_{2}^{(cool)}=-\beta_{3}^{(cool)}$, $\beta_{2}^{(solid)}\approx0$,
 $\beta_{1}^{(solid)}=\beta_{3}^{(solid)}$, $\beta_{2}^{(ms)}>0$, 
 $\beta_{1}^{(ms)}\ge\beta_{3}^{(ms)}\approx0$. In particular, it follows from above formulae
 that the effective cooling-induced anisotropy is uniaxial with the anisotropy axis oriented
 along the tube. In the limit of infinitely thin tube, the axial contribution
 to the shape anisotropy becomes negligible; $\beta_{2}^{(ms)}\gg\beta_{1}^{(ms)}\approx0$. 

Searching for the static solutions to (\ref{LLG}) and performing the micromagnetic simulations
 of tubes, we restrict our considerations to the regimes of solidification-dominated stress and
 cooling-dominated stress. Also, we focus on thin-wall tubes
 taking $\beta_{1}^{(ms)}=\beta_{3}^{(ms)}=0$ in analytical evaluations. Thus, 
 we consider a quasi-2D system with a periodic boundary condition relevant to the tube
 geometry. It is basically in-the-plane magnetized due to the magnetostatics.

\subsection{Tubes with cooling-dominated stress}

According to the above analysis of the anisotropy constants, the cooling-dominated 
 stress produces an axial anisotropy mainly. Therefore, having in mind the aim of obtaining
 the periodic along the tube axis solutions, first, we seek for single-DW solutions assuming
 the domains to be magnetized longitudinally to the wire.
 Thus, the boundary condition ${\rm lim}_{|x|\to\infty}{\bf m}=\pm(M_{s},0,0)$ 
 is satisfied. Using a systematic approach
 of the soliton theory, we look for the equations of motion in the multi-linear form.
 Following \cite{bog80}, we apply the transform 
\begin{eqnarray}
m_{+}=\frac{2M_{s}}{f^{*}/g+g^{*}/f},
\hspace*{2em}
m_{x}=M_{s}\frac{f^{*}/g-g^{*}/f}{f^{*}/g+g^{*}/f},
\label{transform}
\end{eqnarray}
where $m_{\pm}\equiv m_{y}\pm{\rm i}m_{z}$, and we find tri-linear equations of motion for
 the complex functions $g(x,y,z,t)$, $f(x,y,z,t)$ (secondary dynamical variables)
\begin{eqnarray}
-f{\rm i}D_{t}f^{*}\cdot g=f\left[\alpha D_{t}+J(D_{x}^{2}+D_{y}^{2}+D_{z}^{2})
\right]f^{*}\cdot g
\nonumber\\
+Jg^{*}(D_{x}^{2}+D_{y}^{2}+D_{z}^{2})g\cdot g
\nonumber\\
-\frac{2\beta_{1}+\beta_{2}-\beta_{3}}{2}|f|^{2}g
-\frac{\beta_{2}+\beta_{3}}{2}\frac{(y+{\rm i}z)^{2}}{y^{2}+z^{2}}f^{*2}g^{*}\\
-g^{*}{\rm i}D_{t}f^{*}\cdot g=g^{*}\left[\alpha D_{t}-J(D_{x}^{2}+D_{y}^{2}+D_{z}^{2})
\right]f^{*}\cdot g
\nonumber\\
-Jf(D_{x}^{2}+D_{y}^{2}+D_{z}^{2})f^{*}\cdot f^{*}
\nonumber\\
+\frac{2\beta_{1}+\beta_{2}-\beta_{3}}{2}|g|^{2}f^{*}
+\frac{\beta_{2}+\beta_{3}}{2}\frac{(y-{\rm i}z)^{2}}{y^{2}+z^{2}}g^{2}f.
\nonumber
\label{secondary-eq}
\end{eqnarray}
Here $D_{t}$, $D_{x}$, $D_{y}$, $D_{z}$ denote Hirota operators of differentiation 
$D_{x}^{n}b(x,y,z,t)\cdot c(x,y,z,t)\equiv
(\partial/\partial x-\partial/\partial x^{'})^{n}\times\\
\times
b(x,y,z,t)c(x^{'},y^{'},z^{'},t^{'})|_{x=x^{'},y=y^{'},z=z^{'},t=t^{'}}$.

Specific stationary solutions of the single-DW type (for $\beta_{4}=0$) have been identified
 applying the ansatz
\begin{eqnarray}
f=1,\hspace*{2em}g={\rm e}^{kx+{\rm i}\phi+qR\cdot{\rm arctan}(z/y)}
\label{ansatz}
\end{eqnarray}
 to the limit of thin-wall tube $\sqrt{y^{2}+z^{2}}=R$.
 In the case $\beta_{2}+\beta_{3}=0$, denoting $q^{'}\equiv{\rm Re}q$, $q^{''}\equiv{\rm Im}q$, 
 one finds
\begin{eqnarray}
k^{'}k^{''}=-q^{'}q^{''},\hspace*{1ex}
k^{'2}-k^{''2}+q^{'2}-q^{''2}=\frac{K_{a}}{A_{ex}},
\label{solution-conditions}
\end{eqnarray}
 where $K_{a}\equiv M_{s}[\beta_{1}+(\beta_{2}-\beta_{3})/2]/2\gamma$. 
 This case corresponds to the uniaxial effective anisotropy whose axis is 
 oriented along the tube, while the DW solutions are similar to ones studied in the context of 
 relatively-thick (submicrometer) ferromagnetic layers \cite{met05,jan12}. 
 In our effective description of the thin-wall tube, via assuming the domain magnetization to be
 oriented in the tube surface, the magnetostatic contributions to the anisotropy coefficient
 $\beta_{2}^{ms}$ are removed, thus, $\beta_{i}^{ms}=0$ for $i=1,2,3$. Hence,
 with the condition $\beta_{2}+\beta_{3}=0$, we consider the anisotropy due to the 
 cooling-dominated stress in the tube. The conditions of
 periodicity $q^{'}=0$, $q^{''}=l/R$ must be included, where $l$ is integer. 
 The explicit form of the magnetization distribution (written in the cylindrical 
 coordinates; $[x,\rho,\varphi]\equiv[x,\sqrt{y^{2}+z^{2}},{\rm arctan}(z/y)]$) is given by 
\begin{eqnarray}
m_{+}(x,\varphi)&=&M_{s}{\rm e}^{{\rm i}(\phi+q^{''}R\varphi+k^{''}x)}
{\rm sech}\left(k^{'}x+q^{'}R\varphi\right),\nonumber\\
m_{x}(x,\varphi)&=&-M_{s}{\rm tanh}\left(k^{'}x+q^{'}R\varphi\right).
\label{profile1}
\end{eqnarray}

Having found the above single-DW solutions, we modify them when looking for the 
 patterned (multi-domain) periodic textures. We find novel magnetization structures
 just via changing specific hyperbolic functions ${\rm sech}(u)$, ${\rm tanh}(u)$
 in (\ref{profile1}) into the Jacobi elliptic functions;
 ${\rm cn}(u,s^{2})$, ${\rm sn}(u,s^{2})$, $s^{2}\in[0,1)$. This often utilized method
 of finding periodic solutions from soliton solutions is based on common properties of the
 corresponding pairs of the hyperbolic and elliptic Jacobi functions. In particular, 
 describing static DWs with the LLG equation, similarity of the relation
 ${\rm sech}(u){\rm d}^{2}{\rm tanh}(u)/{\rm d}u^{2}-{\rm tanh}(u){\rm d}^{2}{\rm sech}(u)/{\rm d}u^{2}\\=-{\rm sech}(u){\rm tanh}(u)$\hspace*{2em} to \hspace*{2em}
 ${\rm cn}(u,s^{2}){\rm d}^{2}{\rm sn}(u,s^{2})/{\rm d}u^{2}\\-{\rm sn}(u,s^{2}){\rm d}^{2}{\rm cn}(u,s^{2})/{\rm d}u^{2}=-s^{2}{\rm cn}(u,s^{2}){\rm sn}(u,s^{2})$ allows for an easy verification
 of the validity of our method.
 Inserting this way modified magnetization field into (\ref{LLG}), one finds the parameters
 of the periodic solutions to must satisfy similar conditions to (\ref{solution-conditions}).
 Two of the basic types of such solutions are distinguished by the orientation of the DWs parallel
 to the domain magnetization. Such an orientation of the DW relative to the domains is preferred
 by the condition of compensation of the positive and negative magnetic volume
 charges $-\nabla\cdot{\bf m}$, \cite{che07}.

The first one of the distinguished periodic solutions is
\begin{eqnarray}
m_{+}(x,\varphi)&=&M_{s}{\rm e}^{{\rm i}(\phi+q^{''}R\varphi)}
{\rm sn}\left(k^{'}x/s,s^{2}\right)\nonumber\\
m_{x}(x,\varphi)&=&-M_{s}{\rm cn}\left(k^{'}x/s,s^{2}\right),\label{m.a}
\end{eqnarray}
with $k^{'2}+q^{''2}=-K_{a}/A_{ex}$, and $q^{''}R$ to be an integer.
 In the case of $\phi=\pi/2,3\pi/2$, and $s^{2}\approx1$, (\ref{m.a}) relates to the long axis
 of the tube to be magnetically hard. 

Another static solution to (\ref{LLG}) is approximately described with
\begin{eqnarray}
m_{+}(x,\varphi)&=&M_{s}{\rm e}^{{\rm i}\phi}
\left[{\rm cn}\left(\tilde{k}^{''}x/w,w^{2}\right)
\right.\nonumber\\&&\left.
+{\rm i}{\rm sn}\left(\tilde{k}^{''}x/w,w^{2}\right)\right] 
{\rm cn}\left(q^{'}R\varphi/s,s^{2}\right)\nonumber\\
m_{x}(x,\varphi)&=&-M_{s}{\rm sn}\left(q^{'}R\varphi/s,s^{2}\right),
\label{m.c}
\end{eqnarray}
where $q^{'2}-\tilde{k}^{''2}=M_{s}^{2}K_{a}/2A_{ex}$. For the single walled tube,
 $q^{'}2\pi R/s$ is equal to a multiple of the period of the Jacobi functions which
 is the elliptic integral
 $K(s)=\int_{0}^{2\pi}[1-s^{2}{\rm sin}^{2}(\theta)]^{-1/2}{\rm d}\theta$, 
 ($q^{'}2\pi R/s=nK(s)$, where $n$ is integer). In the case $w\neq0$, 
 (\ref{m.c}) satisfies (\ref{LLG}) on the lines $q^{'}R\varphi/s=K(s)h$, 
 where $h=0,1,\ldots,n-1$, and on the lines $\tilde{k}^{''}x/w=K(w)m$,
 where $m=0,\pm1,\pm2,\ldots$. Thus, (\ref{LLG}) is satisfied on lines of a net that
 covers the tube surface. Therefore, we consider (\ref{m.c}) to be an approximate
 representation of a possible texture. The intersections of the lines of both series
 (the net sites) coincide with the centers of vortices or antivortices inherent in the texture. 

In the $w=0$ limit case of (\ref{m.c}), that is reached via exchanging $\tilde{k}^{''}/w$
 into $k^{''}$, thus, exchanging
 ${\rm cn}(\tilde{k}^{''}x/w,w^{2})+{\rm i}{\rm sn}(\tilde{k}^{''}x/w,w^{2})$
 in the first of equations (\ref{m.c}) into ${\rm e}^{{\rm i}k^{''}x}$, one obtains a strict
 solution to (\ref{LLG}). Its parameters satisfy $q^{'2}-k^{''2}=K_{a}/A_{ex}$.

\subsection{Tubes with solidification-dominated stress}

According to previous section considerations, in the thin-wall tube, the solidification
 contribution to the constant of the circumferential anisotropy is equal to that
 of the axial anisotropy;
 $\beta_{1}^{(solid)}=\beta_{3}^{(solid)}$, and the contribution to the radial anisotropy 
 constant $\beta_{2}^{(solid)}$ is equal to zero. However, such a two-axis
 anisotropy is equivalent to the radial anisotropy whose type (easy-axis or hard-axis)
 is determined by the sign of the magnetostriction constant. Thus, the consequence
 of the solidification stress is the enhancement or attenuation of the magnetostatically-induced
 radial anisotropy of the hard-axis type. Provided the summary radial hard-axis anisotropy
 is sufficiently strong, the efficient way to describe the magnetic ordering is applying the XY
 model on the curved surface (the tube surface) spanned on the axial (X) and circumferential (Y)
 directions. Detailed justification of that approach to the ordering in thin soft-ferromagnetic
 film has been given elsewhere \cite{jan13}. We write the magnetization in the cylindrical
 coordinates and assume $m_{\rho}=0$
 and $m_{x}+{\rm i}m_{\varphi}=M_{s}{\rm e}^{{\rm i}(\eta+\theta)}$ with $\eta={\rm const}$.
 Insertion of the above magnetization into (\ref{LLG}), in the static case,
 leads to the Laplace equation 
\begin{eqnarray}
\frac{\partial^{2}\theta}{\partial x^{2}}
+\frac{1}{R^{2}}\frac{\partial^{2}\theta}{\partial\varphi^{2}}=0.
\end{eqnarray}
Utilizing its DW solution from \cite{jan13}, one arrives at 
\begin{eqnarray}
m_{x}=M_{s}\frac{{\rm cos}(\eta)b-{\rm sin}(\eta)a}{\sqrt{a^{2}+b^{2}}},\nonumber\\
m_{\varphi}=M_{s}\frac{{\rm sin}(\eta)b+{\rm cos}(\eta)a}{\sqrt{a^{2}+b^{2}}},
\label{transform-XY}
\end{eqnarray}
where $a\equiv{\rm sin}(k^{''}x+q^{''}R\varphi)$, $b\equiv{\rm sinh}(k^{'}x+q^{'}R\varphi)$,
 and the conditions $k^{'2}-k^{''2}+q^{'2}-q^{''2}=0$, $k^{'}k^{''}=-q^{'}q^{''}$ are satisfied.
 This single-DW solution cannot satisfy the periodic boundary condition, 
 ${\bf m}(x,\varphi)\neq{\bf m}(x,\varphi+2\pi)$ for $q^{'}\neq 0$. Therefore, it is not 
 relevant to the single-walled tube. Exchanging $b\equiv{\rm sinh}(k^{'}x+q^{'}R\varphi)$
 in (\ref{transform-XY}) into
 $b\equiv{\rm sn}(k^{'}x/s+q^{'}R\varphi/s,s^{2})/{\rm dn}(k^{'}x/s+q^{'}R\varphi/s,s^{2})$,
 for $|s|$ very close to 1, one obtains an approximate solution to the XY model 
 that is periodic in the axial direction and in the circumferential direction provided
 $q^{''}R=l$ and $q^{'}2\pi R\approx nK(s)$. Here, $n$ and $l$ are integers. That solution 
 represents a system of many parallel DWs.
 When ${\rm tan}(\eta)=q^{'}/k^{'}$, the DWs are parallel to the magnetization of the domains.

The direction of the DW relative to the tube axes follows from a competition between  
 axial (shape) anisotropy of the tube and the anisotropy due
 to the direction of the material deposition. Therefore, it is dependent of the length
 of the tube. We include both anisotropies into the reduced (XY) model just via the boundary
 condition on the direction of the domain magnetization, the angle $\eta$, similar
 to the description of DWs in soft-magnetic nanostripes \cite{jan13,tch05}. However, unlike
 in the nanostripes whose edges strongly affect the DW texture (via stabilizing singularities;
 antivortices), in the case of the infinite planar layers or tube walls, the (anti)vortex
 containing DWs are not preferable for sufficiently thin systems while Neel DWs are \cite{met05,mid63}.
 Those DWs are not described with the XY model.   

Besides the geometry of the tube (thickness of its wall), the applicability of the XY model is
 dependent on the ratio of the magnetostatic exchange
 length $l_{ms}\equiv\sqrt{2A_{ex}/\mu_{0}M_{s}^{2}}$ to the exchange length of the
 solidification-stress-induced (helical) anisotropy
 $l_{K}\equiv\sqrt{A_{ex}/K_{h}^{(solid)}}$,
 where $K_{h}^{(solid)}\equiv M_{s}\beta_{4}^{(solid)}/2\gamma$.
 The magnetostatic field and the helical-anisotropy field compete in the area of the DW. 
 When, $l_{ms}$ is significantly smaller than $l_{K}$, (the magnetostatic field is stronger
 than the helical-anisotropy field), which enforces the ordering to be completely confined
 to the tube surface, (the Neel-DWs are preferable instead of the cross-tie DWs). 
 In the parameter range of $l_{ms}$ close to $l_{K}$, both fields can cancel each other,
 which makes the XY model applicable. In the former case, ($l_{ms}\ll l_{K}$), the magnetostatic 
 exchange length corresponds to DW width and the magnetostatic field in the DW area can be modeled
 with a $M_{s}\beta_{4}^{(ms)}\equiv2\gamma K_{h}^{(ms)}\sim\gamma\mu_{0}M_{s}^{2}$ contribution
 to the anisotropy constant in (\ref{LLG}) while the DW magnetization can be described with
 direct solutions to the LLG equation. 
 
For $\beta_{2}+\beta_{3}=0$, in the limit of weak shape (axial)
 anisotropy; $|\beta_{1}|\ll\beta_{4}$, the relevant DW texture is described with a rotated
 in the $x-\varphi$ plane function (\ref{profile1});
\begin{eqnarray}
m_{x}(x,\varphi)&=&-M_{s}{\rm cos}(\eta)c-M_{s}{\rm sin}(\eta)d,\nonumber\\
m_{\rho}(x,\varphi)&=&M_{s}e,\label{transform-XYX}\\
m_{\varphi}(x,\varphi)&=&-M_{s}{\rm sin}(\eta)c+M_{s}{\rm cos}(\eta)d,
\nonumber
\end{eqnarray}
where
\begin{eqnarray}
c&\equiv&-{\rm tanh}(k^{'}x+q^{'}R\varphi),\nonumber\\
d&\equiv&[{\rm cos}(\varphi){\rm cos}(\phi+q^{''}R\varphi+k^{''}x)
+{\rm sin}(\varphi)\nonumber\\
&&\times{\rm sin}(\phi+q^{''}R\varphi+k^{''}x)]
{\rm sech}(k^{'}x+q^{'}R\varphi),\label{transform-XYZ}\\
e&\equiv&[-{\rm sin}(\varphi){\rm cos}(\phi+q^{''}R\varphi+k^{''}x)
+{\rm cos}(\varphi)\nonumber\\
&&\times{\rm sin}(\phi+q^{''}R\varphi+k^{''}x)]
{\rm sech}(k^{'}x+q^{'}R\varphi),\nonumber
\end{eqnarray}
and
\begin{eqnarray}
k^{'}k^{''}=-q^{'}q^{''},\label{transform-sup1}\\
k^{'2}-k^{''2}+q^{'2}-q^{''2}=\frac{K_{h}}{A_{ex}}.
\label{transform-sup2}
\end{eqnarray}
Here, $K_{h}=K_{h}^{(ms)}+K_{h}^{(solid)}$, and $K_{h}^{(ms)}K_{h}^{(solid)}<0$.
We expect the DW to be parallel to the domain magnetization, thus; ${\rm tan}(\eta)=q^{'}/k^{'}$.
 Similar to the formula (\ref{transform-XY}),
 the formula (\ref{transform-XYZ}) is not applicable  
 to single-walled tube because one cannot implement the periodic boundary condition;
 ${\bf m}(x,\varphi)\neq{\bf m}(x,\varphi+2\pi)$. In order to do it,
 we find another solution to (\ref{LLG}) via exchanging in (\ref{transform-XYZ})
 the hyperbolic functions into elliptic ones;
 ${\rm sech}(k^{'}x+q^{'}R\varphi)\to{\rm sn}(k^{'}x/s+q^{'}R\varphi/s,s^{2})$ and
 ${\rm tanh}(k^{'}x+q^{'}R\varphi)\to{\rm cn}(k^{'}x/s+q^{'}R\varphi/s,s^{2})$, as well as
 transforming the condition (\ref{transform-sup2}) into
\begin{eqnarray}
k^{'2}+k^{''2}+q^{'2}+q^{''2}=-\frac{K_{h}}{A_{ex}}.
\label{transform-sup3}
\end{eqnarray}
While the formulae (\ref{transform-XYX})-(\ref{transform-sup2}) are capable to describe cross-tie
 DWs even in the case $\beta_{4}\neq0$, with $|k^{'}|=|q^{'}|$ and $|q^{''}|=|k^{''}|$
 for instance \cite{jan12}, the periodic boundary condition restricts values
 of $q^{'}$ and $q^{''}$ to discrete sets. Applied together with (\ref{transform-sup1}),
 (\ref{transform-sup3}), this condition drives $q^{''}=k^{''}=0$. Therefore, those
 cross-tie DWs cannot be stabilized in the single-walled tube. Hence, unlike the DWs
 of the XY model, the present DW does not contain	vortices nor antivortices, which provides
 basic distinction between the multi-domain textures	relevant to different regimes of the
 description of the tubes with solidification-dominated stress.

\begin{figure} 
\unitlength 1mm
\begin{center}
\begin{picture}(175,100)
\put(0,-2){\resizebox{88mm}{!}{\includegraphics{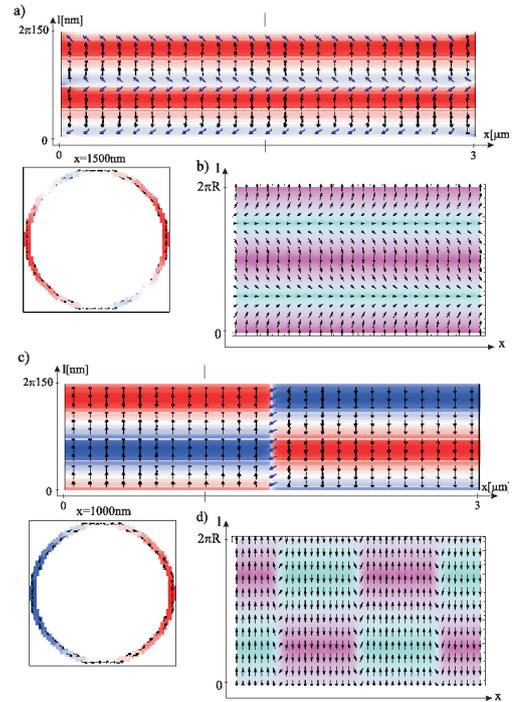}}}
\end{picture}
\end{center}
\caption{The textures established in the Co tube (of 300nm diameter, 3$\mu$m length,
and 20nm wall thickness) with micromagnetic simulations, for the initial magnetization of the tube
$[M_{s}/\sqrt{2},M_{s}/\sqrt{2},0]$ (a), $[M_{s},0,0]$ (c). For comparison to numerical results,
analytical expressions for the tube magnetization are plotted; the texture relevant to the $w=0$
limit of expression (\ref{m.c}) and $k^{''}=0$, $s\approx 0$, $|q^{'}|R\approx 1$ (b), and
the texture relevant to (\ref{m.a}) for $|s|=1-6\cdot10^{-11}$, $|q^{''}|R=1$ (d).
In a) and c), the constant of (stress-induced) uniaxial anisotropy (of the [1,0,0] hard direction) is $K_{a}^{(cool)}=-1.5\cdot10^{5}$J/m$^{3}$. Here and below, $l\equiv R\varphi$ denotes the circumferential coordinate while area colors and their intensities
indicate sign and value of $m_{y}$ component of the magnetization. Arrow colors
in the simulation-obtained plots indicate sign and value of $m_{x}$.}
\end{figure} 

\begin{figure} 
\unitlength 1mm
\begin{center}
\begin{picture}(175,88)
\put(0,-2){\resizebox{88mm}{!}{\includegraphics{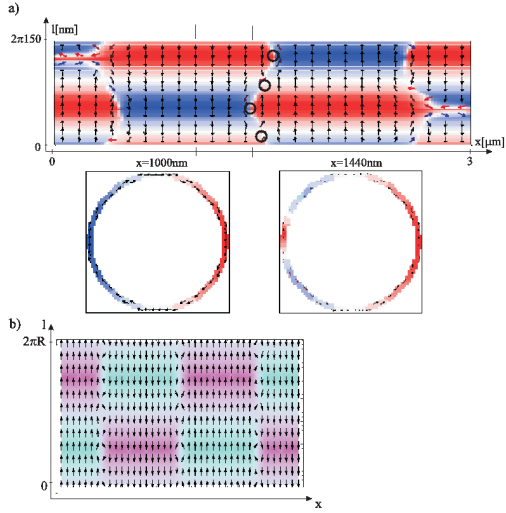}}}
\end{picture}
\end{center}
\caption{The texture established in the Co tube (of 300nm diameter, 3$\mu$m length,
and 20nm wall thickness) with the initial magnetization of the tube $[0,M_{s},0]$ (a).
Point singularities (vortices and antivortices) in the central DW are indicated by circles.
For comparison to numerical results, analytical expression for the tube magnetization;
the textures relevant to (\ref{m.a}) for $|s|=1-6\cdot10^{-11}$ and $|q^{''}|R=4$ is plotted (b).
The material parameters for a) are the same as in Fig. 1. The area (arrow) color and its intensity indicate sign and value of $m_{y}$ ($m_{x}$) component of the magnetization, similar as in Fig. 1.}
\end{figure}
 
\section{Micromagnetic simulations}

We have performed a series of micromagnetic simulations of the relaxation 
 of thin-wall ferromagnetic tubes in order to verify the analytically obtained
 magnetization distributions. Simulating the tubes with the cooling-dominated stress,
 we apply the material parameters of cobalt. In \cite{mul09}, the cooling stress has been 
 directly shown to can significantly overcome the solidification stress in the rolled-up Co
 nanomembrane, which results in qualitative changes in the hysteresis loops of the tubes
 with temperature. Dealing with the 
 solidification-dominated stress, we include the parameters of Permalloy additionally,
 in order to examine systems of significantly different ratio of the magnetostatic exchange
 length $l_{ms}$ to the exchange length of the stress-induced (helical) anisotropy $l_{K}$.
 Notice that the magnetostatic exchange lengths of Co and Py are the same ($l_{ms}=5.2$nm)
 while their exchange stiffnesses $A_{ex}$ are very different.
 The techniques of rolling-up the nanomembranes of these magnetic materials
 are well developed \cite{mul09,str13}, so as related nickel
 nanomembranes \cite{str14,min12,ruf12}.

Due to the computational complexity, the diameters of the simulated
 structures are limited to submicrometer sizes. We applied OOMMF package \cite{oommf}.
 The discretization size of the tube grid is 5nm in the YZ plane and 10nm
 in the X direction. The Gilbert damping constant is $\alpha=0.5$.

\begin{figure} 
\unitlength 1mm
\begin{center}
\begin{picture}(175,74)
\put(0,-2){\resizebox{88mm}{!}{\includegraphics{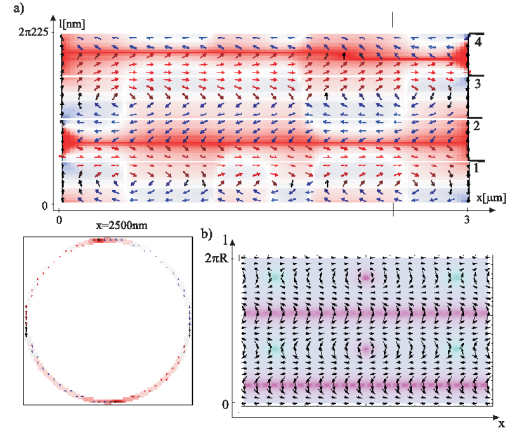}}}
\end{picture}
\end{center}
\caption{The texture established in the Co tube of 450nm diameter, 3$\mu$m length,
and 20nm wall thickness with micromagnetic simulations, for the initial magnetization of the tube
$[0,M_{s},0]$ (a). The constant of (stress-induced) uniaxial anisotropy (of the [1,0,0] hard
direction) is $K_{a}^{(cool)}=-0.63\cdot10^{5}$J/m$^{3}$. Meaning of colors is the same
as in Figs. 1-2. Four sections of the tube wall are indicated with numbers 1-4.
For comparison to numerical results, relevant analytical expression for the tube magnetization is plotted (b). Sections $\pi R/2<l<\pi R$ or $3\pi R/2<l<2\pi R$; [relevant to sections 1 and 3 of a)]
are described with (\ref{m.c}) for $q^{'2}\approx\tilde{k}^{''2}$,
$|s|=1-6\cdot10^{-3}$, $|w|=1-6\cdot10^{-6}$,
while others [relevant to sections 2 and 4 of a)] with the $w=0$ limit of (\ref{m.c}) and $k^{''}=0$, $s\approx 0$, $|q^{'}|R\approx 2$.}
\end{figure}

\subsection{Tubes with cooling-dominated stress}

In the simulations of the Co tubes, the following material constants are used;
 the saturation magnetization $M_{s}=1.4\cdot10^{6}$A/m, the exchange
 stiffness $A_{ex}=3.3\cdot10^{-11}$J/m. Assuming the stress to be cooling-dominated,
 the stress-induced axial anisotropy of the easy-plane type is included
 with an anisotropy constant $K_{a}^{(cool)}\sim(-1)\cdot10^{4}\div(-1)\cdot10^{5}$J/m$^{3}$,
 [$K_{a}^{(cool)}\sim M_{s}(\beta_{1}^{(cool)}-\beta_{3}^{(cool)})/2\gamma$]. In \cite{mul09},
 a Co tube has been reported to possess the axial anisotropy relevant
 to $K_{a}\sim10^{3}$J/m$^{3}$
 for the diameter of several micrometers while cooling of the tube to result in decrease
 of the anisotropy constant to $K_{a}\sim(-1)\cdot10^{4}$J/m$^{3}$, (thus, in the change of the
 easy-axis into the hard-axis). Decreasing the tube diameter down to 
 several hundreds of nanometers, the absolute value of the density of energy
 of the stress-driven anisotropy is expected to increase. The Co tubes of 300nm diameter,
 20nm wall thickness and 3$\mu$m length, and of the arbitrarily chosen anisotropy constant 
 $K_{a}^{(cool)}=-1.5\cdot10^{5}$J/m$^{3}$ have been simulated with different initial conditions. 
 The calculations begun with the homogeneous magnetization
 ${\bf m}(t=0)=[M_{s}/\sqrt{2},M_{s}/\sqrt{2},0]$ and finished with the stable configuration of Fig. 1a.
 Correspondingly, initialized with the longitudinal magnetization ${\bf m}(t=0)=[M_{s},0,0]$ 
 simulation ended with the texture of Fig. 1c, while the perpendicular
 magnetization ${\bf m}(t=0)=[0,M_{s},0]$ evolved to that of Fig. 2a. One sees that there are 
 many stable (metastable) states of the tube magnetization which realize under 
 specific initial conditions. They correspond to different stationary solutions
 to the LLG equation.
 
In particular; Fig. 1a corresponds to a field state (\ref{m.c})
 with $w=0$, $k^{''}=0$, $s^{2}\approx 0$, $q^{'}2\pi R/s=\pm K(s)$, $\phi=\pi/2$ that 
 is plotted in Fig. 1b.
 Hence, the magnetization is pointed in the XZ plane. The texture in Fig. 1b is found via matching
 solutions of positive and negative $q^{'}$ on the lines $\varphi=0$ and $\varphi=\pi$.

The texture of circumferentially magnetized domains in Figs. 1c corresponds to (\ref{m.a}) 
 with $|q^{''}|R=1$, $s^{2}\approx 1$, $\phi=\pi/2$ (plotted in Fig. 1d). Besides the vicinities of
 the wire ends, the domains of Fig. 2a are similar to those of Fig. 2b. The plot of Fig 2b
 has been obtained similar to that of Fig. 1d while taking $|q^{''}|R=4$. The centers of 
 vortices and antivortices which are contained in the structure of the DWs lie on
 the lines $\varphi=n\pi/4$, where $n=0,1,\ldots,7$. From Fig. 2b
 the relevant DWs are seen to be of the cross-tie type similar to the DWs of Fig. 2a.
 Note that a denser bamboo-like domain structure has been established
 within simulations of a similar to ours nanotube of a ferromagnet with considerably 
 larger magnetostatic exchange length than that of Co \cite{uso11}.
 We mention that microscopic mechanism responsible for appearance of the hard-axis anisotropy
 (positive axial stress), thus, for creation of the bamboo-like textures in glass-coated wires
 of the materials with negative magnetostriction is not completely clear to date \cite{uso14},
 however, it is known to be activated during the cooling process. In the thin-wall tubes
 (rolled-up nanomembranes) it seems to be especially efficient because of smaller solidification
 stresses than in the wires.
 A general model of a similar DW to that of Fig. 1c while including effects of
 finite thickness of the tube wall (\cite{che12}) has been formulated in \cite{che15}. 
 A cross-tie DW similar to those in Fig. 2a and Fig. 2b has been simulated in \cite{bet08}. 

In Fig. 3a, the relaxed magnetization of the tube of 450nm
 diameter, 20nm wall thickness and 3$\mu$m length, and initially
 magnetized perpendicular to the long axis (${\bf m}(t=0)=[0,M_{s},0]$) is visualized.
 Without detailed estimation of the stress, we include the arbitrary value of the constant
 of the solidification-stress induced anisotropy $K_{a}^{(cool)}=-0.63\cdot10^{5}$J/m$^{3}$
 that is reduced compared to that applied
 in the simulations of the tubes of 300nm diameter. The reduction in the stress 
 due to the reduction of the tube-surface curvature is expected and it results in a decrease
 of the  effective anisotropy constant. The texture is found to contain vortices
 and antivortices whose centers are magnetized along the axis 
 of the initial magnetization of the tube. 

\begin{figure} 
\unitlength 1mm
\begin{center}
\begin{picture}(175,51)
\put(0,-5){\resizebox{88mm}{!}{\includegraphics{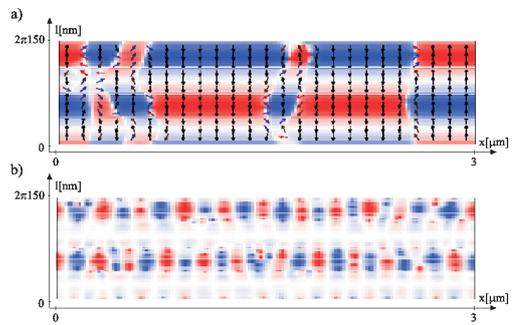}}}
\end{picture}
\end{center}
\caption{The textures established in the Co tube of 300nm diameter, 
3$\mu$m length, 20nm wall thickness (a) and 5nm wall thickness (b)
with the random initial orientations of the magnetization.
The material parameters, the reference frame and meaning
of colors are the same as in Figs. 1-2. In b), the arrows are omitted 
for better visibility.}
\end{figure}

The texture from Fig. 3a can be quite well described locally with the static
 solution (\ref{m.c}) to the LLG equation.
 In the sections of the tube; $0<\varphi<\pi/2$
 and $\pi<\varphi<3\pi/2$ in Fig. 3b, (these sections correspond to sections 2 and 4 in Fig. 3a),
 $w=0$ and $k^{''}=0$, thus, $q^{'2}=(\beta_{1}-\beta_{3})/J$. Since $\beta_{1}>\beta_{3}$,
 the magnetostatic contribution to the anisotropy constant oversizes the magnetostrictive
 contribution leading to the easy-axis. In the remaining (vortex-containing) sections of the tube 
 (which correspond to sections 1 and 3 in Fig. 3), there are no well-defined domains, thus,
 there is no easy direction of the magnetization in the plane of the tube surface.
 There, the magnetostatic (shape) contribution to the the anisotropy constant has to be
 compensated by the stress contribution, thus, $q^{'2}\approx\tilde{k}^{''2}$.
 In Fig. 3b, in each of four $h\pi/2<\varphi<(h+1)\pi/2$ sections, ($h=0,1,2,3$); 
 $q^{'}2\pi R/s=2K(s)$.

Finally, we have simulated the relaxation from a disordered initial state (a random
 magnetization orientation). In the system of Figs. 1-2, (the Co tube of 300nm diameter,
 3μm length, and 20nm wall thickness), the resulting state is formed of several
 circumferentially-magnetized domains which are of different (irregular)
 lengths (Fig. 4a). However, for the 5nm tube-wall thickness, the
 domains of circumferential magnetization are very short while the texture 
 becomes almost periodic (Fig. 4b). That state of a tube with an ultra-thin 
 wall resembles a periodic texture found in Ref. \cite{uso11}
 within 2D micromagnetic simulations.

\subsection{Tubes with solidification-dominated stress}

\begin{figure} 
\unitlength 1mm
\begin{center}
\begin{picture}(175,55)
\put(0,-5){\resizebox{88mm}{!}{\includegraphics{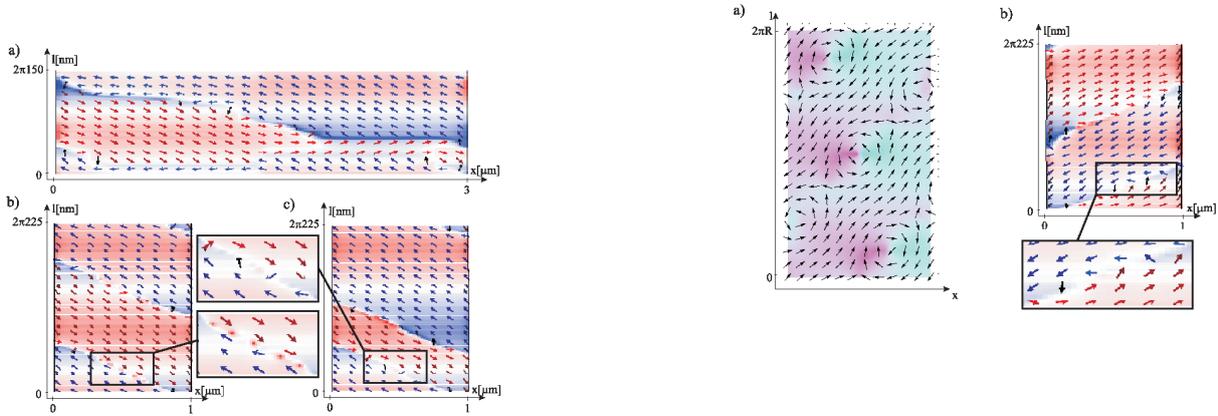}}}
\end{picture}
\end{center}
\caption{The stable texture established in the Py tube: a) of 300nm diameter, 3$\mu$m length,
and 20nm wall thickness, and helical anisotropy
constant $K_{h}^{(solid)}=-0.3\cdot10^{5}$J/m$^{3}$, with the initial magnetization $[0,M_{s},0]$;
b) and c) of 450nm diameter, 1$\mu$m length, and 20nm wall
thickness, $K_{h}^{(solid)}=-1.0\cdot10^{5}$J/m$^{3}$, with the initial magnetization $[0,M_{s},0]$ and $[M_{s}/\sqrt{2},M_{s}/\sqrt{2},0]$ respectively.
Meaning of colors is the same as in Figs. 1-2. In the inset of b), a chain of vortex (antivortex)
 cores of the same polarity is seen inside the DW. In the inset of c), a single isolated 
 defect (a vortex without any clear polarity) is seen in the DW.}
\end{figure}

The tubes of Py have been simulated with the following material constants:
 $M_{s}=0.86\cdot10^{6}$A/m, $A_{ex}=1.3\cdot10^{-11}$J/m. In the regime of the 
 solidification-dominated stress, visible effects are expected to result from 
 specific orientations of the easy direction in the tube wall. This anisotropy 
 is induced by the material deposition at the stage of the production of the 
 magnetic layer. We simulate the Py tube with a resulting helical anisotropy
 whose axis is deviated by $\pi/4$ angle from the long axis of the stripe
 in the tube wall (directed along $[\sqrt{y^{2}+z^{2}},-z,y]$ vector). 
 If this anisotropy is sufficiently strong, it drives the spiral orientation of 
 the magnetization inside the domains. We consider a constant the helical anisotropy
 in the range $K_{h}^{(solid)}\sim(-1)\cdot10^{4}\div(-1)\cdot10^{5}$J/m$^{3}$.
 Notice that in-the-plane anisotropy of the constant (absolute value) close to $10^{4}$J/m$^{3}$
 has been reported for flat and rolled-up nanomembranes of Py  \cite{str14}, and for
 Co film \cite{dij01}. 

In Fig. 5a, the resulting ordering of Py tube of 300nm diameter, 20nm wall thickness,
 3$\mu$m length and of $K_{h}^{(solid)}=-0.3\cdot10^{5}$J/m$^{3}$ is visualized. 
 In this quite elongated and of weak anisotropy tube, the competition between the 
 shape and structural anisotropy creates a 'frustration' of the magnetic moments. They
 orient along the tube axis at some parts of the tube while, at other parts, they tend to
 point onto the spiral line. In consequence, there are no well separated (by DWs)
 ferromagnetic domains in the texture of Fig. 5a.    
 
\begin{figure} 
\unitlength 1mm
\begin{center}
\begin{picture}(175,45)
\put(0,-5){\resizebox{88mm}{!}{\includegraphics{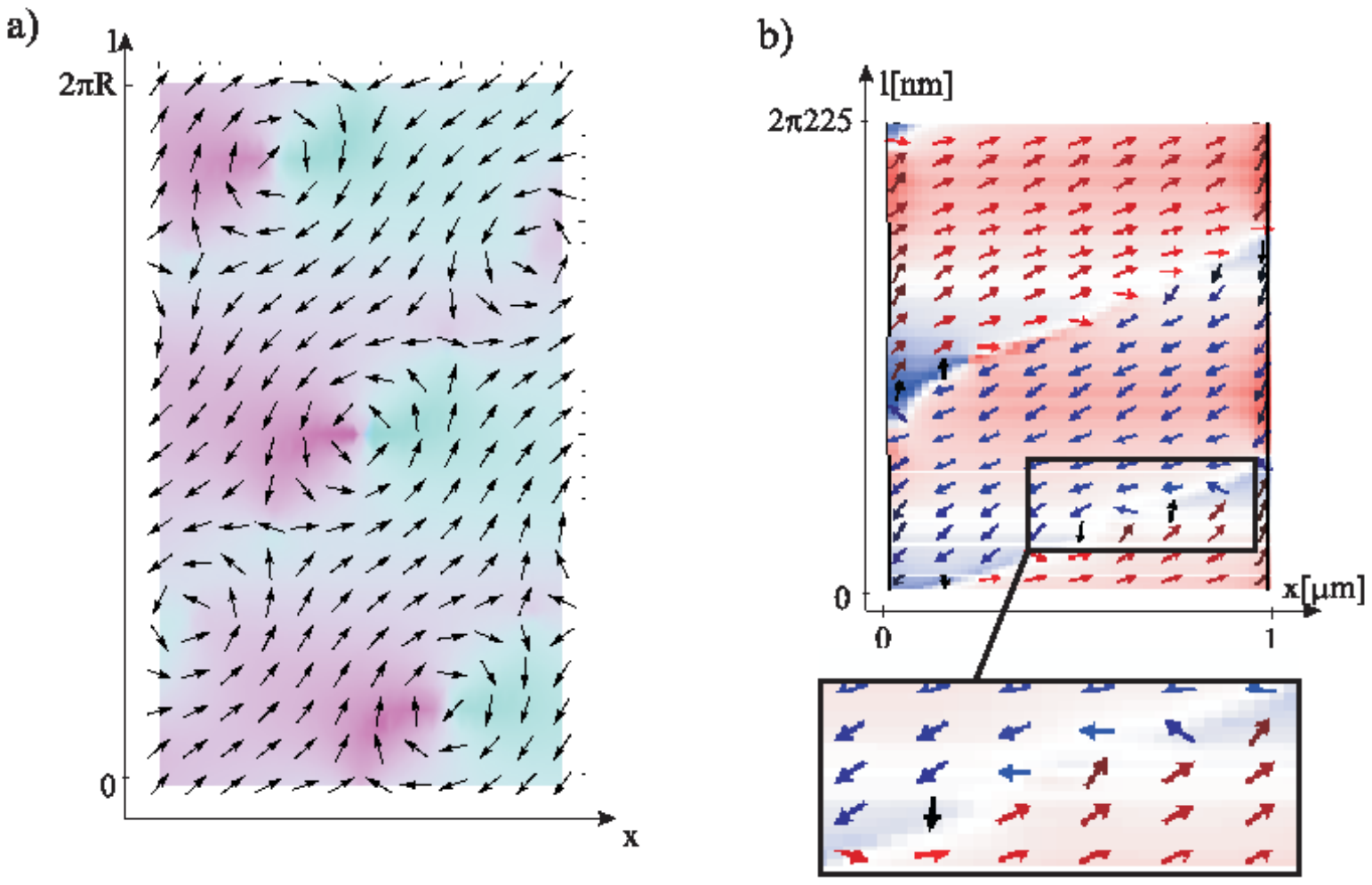}}}
\end{picture}
\end{center}
\caption{Plot of a spatially-periodic solution to the XY model on the tube surface (a).
The stable texture (b) established in the Co tube of 450nm diameter, 1$\mu$m length, and 20nm wall thickness, and helical anisotropy constant $K_{h}^{(solid)}=-1.0\cdot10^{5}$J/m$^{3}$, with the initial magnetization $[0,M_{s},0]$. Meaning of colors is the same as in Figs. 1-2.}
\end{figure}

The next two visualizations (Figs. 5b, 5c) represent the static ordering of the Py tube of 450nm
 diameter, 20nm wall thickness, 1$\mu$m length and of $K_{h}^{(solid)}=-1.0\cdot10^{5}$J/m$^{3}$.
 Its longitudinal (shape) anisotropy is reduced compared to the structure of Fig. 6a 
 due to the small length of the system and the ordering is dominated by the material
 anisotropy. The well defined domains of spiral ordering are shown
 to be separated by DWs which are parallel to the domain magnetization. 
 Similar to the simulations of the tubes with cooling-dominated stress,
 the final state of the magnetization relaxation is dependent on the initial state. For instance,
 in Figs. 5b and 5c, the stable textures obtained with the homogeneous initial
 states ${\bf m}(t=0)=[0,M_{s},0]$ and ${\bf m}(t=0)=[M_{s}/\sqrt{2},M_{s}/\sqrt{2},0]$, respectively,
 are presented. The stabilized textures differ not only in the position of the DWs, while also
 in the DW structure.  

The ordering in Fig. 5b coincides with the analytically predicted DW textures for the XY model.
 The relevant DW-containing texture described with (\ref{transform-XY}) for $|\eta|=\pi/4$ is 
 plotted in Fig 6a. Such a DW contains alternating vortices and antivortices whose polarity
 is not determined by the model. In the inset of Fig. 5b, one sees all singularities of the DW
 to can be of the same polarity. This is not obtainable
 from direct cross-tie-DW solutions to the LLG equation of the one-axis ferromagnet
 in 2D whose vortex and antivortex polarities are opposite \cite{met05,jan12}.
 In contrast, the DWs in Fig. 5c are not of the cross-tie structure. 

In Fig. 6b, the result of the simulation of the cobalt tube of the same sizes as the tube 
 in Fig. 5b with similar initial condition ${\bf m}(t=0)=[0,M_{s},0]$ is presented. In the Co tube, 
 very different (Neel-like) DWs from (cross-tie) DWs of the Py tube are found because
 of bigger exchange length of the helical anisotropy $l_{K}$ for Co
 ($l_{K}=18.2$nm) than for Py ($l_{K}=11.4$nm). This shows the inapplicability
 of the XY model to the Co tube contrary to the Py tube.
 Noticeably, the applicability of the XY model is dependent on the initial
 conditions as well, which follows from the difference in the DW structures
 in Fig. 5b and Fig. 5c.

Due to finite length of the tube, the DWs in Figs. 5c,6b are not perfect Neel structures while
 contain isolated vortices which are not predicted by our analytical description of the infinite
 tubes. We note that the polarities of large-core vortices (antivortices) present in the textures
 of Figs. 3,5c,6b are not determined by the simulations because of internal domain structure
 of the vortex and antivortex \cite{you06}.

Studying the relaxation of tubes of different wall thicknesses from the state of
 randomly-distributed orientations of the magnetization, for a wide range of the material
 parameters and tube sizes, with $\pi/4$ angle of the easy-axis deviation
 from the tube long axis, we find the resulting texture to be
 a helical single-domain state. In particular, for the systems of Fig. 5 and Fig. 6.
 The magnetization is oriented along the easy direction in the tube surface. However,
 the initial stage of the relaxation process in thin-wall tubes is found to be the formation
 of a quasi-periodic state that contains long oppositely-magnetized helical domains. Formed
 with the random initial condition, those striped domains appear not to
 be stable and slowly collapse into a mono-domain (helically-magnetized) state.
 We mention that the observation of striped helical domains in the rolled-up nanomembranes
 has been reported however \cite{str14}. Perhaps, upon being created,
 they are stabilized by tube imperfections; structural defects
 or a jump on the tube surface that corresponds the edge of the rolled-up sheet
 of a ferromagnetic material.
  
\section{Conclusions}

A series of analytical static solutions to LLG equation for an effective model of the ferromagnetic
 thin-wall tube has been found analyzing different parameter regimes. The applicability 
 of the solutions of our model (that is based on the substitution of the magnetostatic field
 in the tube with a contribution to an effective anisotropy field) has been verified using
 the micromagnetic simulations. Including one of two (axial or helical) types of the effective
 anisotropy, the model appeared to be capable to describe basic features of the ordering
 in ferromagnetic thin-wall tubes.

It has been established that main characteristics of the magnetic textures (the number, size, and 
 shape of the domains) are governed by the stress-induced anisotropy. Since the stress
 in the tube is sensitive to details of the tube fabrication, the ordering is not universal
 result of the material and size parameters. Moreover, a strong dependence of the ordering on
 initial state of the structure formation process is noticed. In spite of this, there is a 
 quite limited basis of possible magnetization textures of the tube and factors
 influencing their choice are identified. 

Details of textures (the structure of the DWs) are dependent on the ratio of 
 the magnetostatic exchange length to the exchange length of the stress-driven anisotropy.
 This makes a problem for efficiently simulating
 tubes (rolled-up nanomembranes) of the sub-$\mu$m- and $\mu$m-radii due to large
 computational resources needed when including the magnetostatic (dipole) interactions.
 To date, in order to avoid this problem, any approach to
 the micromagnetic simulations of such systems of a large-number of the magnetic moments
 is based on oversizing the length of the grid discretization over the magnetostatic
 exchange length \cite{sto12,gaw15}. This can result in overseeing some texture details. 
 Our analytical model enables inclusion of them into the tube description.

Finally, we mention that we have not considered nano- nor microtubes (rolled-up nanomembranes)
 with perpendicular magnetic anisotropy. In those hard-magnetic systems a dense structure
 of radially magnetized domains is spontaneously created, e.g. in rolled-up Co/Pt and Co/Pd
 layers \cite{zar14,str15}. It is similar to that observed in the outer shell of the amorphous
 or polycrystalline glass-coated microwires with positive magnetostriction constant as a result
 of the competition between the stress-driven radial easy-axis anisotropy
 and the surface (magnetostatically-induced) radial anisotropy of the hard-axis type \cite{chi96,kab05}.
 Since the layers with perpendicular magnetic anisotropy are ultra-thin, 3D micromagnetic 
 simulations of those tubes are challenging, (they require the grid discretization with
 very small cells), even for nanometer-sized radii of the tubes.
 On the other hand, unlike in the microwires with positive magnetostriction,
 the surfaces of the DWs in such tubes are extremely small, thus, pinning of them
 to structure defects is very strong. This makes the magnetic ordering irregular (spatially
 aperiodic) while dominated by structural properties.

\end{document}